# Quantum wave equations and non-radiating electromagnetic sources

Mark P Davidson
Spectel Research Corp., Palo Alto, CA  USA

**Abstract**
A connection between classical non-radiating sources and free-particle wave equations in quantum mechanics is rigorously made.  It is proven that free-particle wave equations for all spins have currents which can be defined which are non-radiating electromagnetic sources.  It is also proven that the advanced and retarded fields are exactly equal for these sources.  Implications of these results are discussed.

## 1. Introduction

Non-radiating electromagnetic sources have a long history in physics with practical applications in plasma physics, optics, electrical engineering, inverse scattering, and with fundamental importance to a number of physical phenomena.  The earliest commonly cited references are [1-5].   The most general form for such current distributions is not known.  There has been an effort to describe elementary particles as extended objects of this type [3-8].  And assorted other applications may be found in [9-21].

It has recently been shown that any solution to the free particle Schrödinger's equation has probability density and current which are classical non-radiating sources even though they vary with both space and time [22]. This result was also found to be true for the Klein-Gordon equation so long as it was restricted to positive (or negative) frequencies. Here we make these arguments more rigorously and extend them to the Dirac equation and to other more general higher spin wave equations. It is also argued that the perfect linearity of Schrodinger's equation as tested in neutron diffraction experiments [23] may be a result of  the fact that nonlinear terms would lead to some radiation in the above context.  The theory of non-radiating sources is also closely related to the electrodynamics of Wheeler and Feynman [24-26].  It is further shown that for all the non-radiating wave equations, the retarded and advanced fields are exactly equal.

It is found that although the Dirac current radiates classically, a non-radiating current can be naturally defined if one first makes a Foldy-Wouthuysen transformation [27] of the Dirac wave equation.  This technique applies to higher spin wave equations as well by using generalizations of the Foldy-Wouthuysen representation [28, 29].

Quantum field theorists may not find these results surprising, as there are many similarities between the classical radiation theory and the quantum radiation theory, and of course quantum free particles do not radiate in quantum electrodynamics (QED). Nevertheless, studying the problem rigorously and entirely within the framework of classical electromagnetism is interesting because it sheds light on the non-radiating source research going on in many different fields, can lead to formerly unknown non-radiating sources which may have practical value, and may have value in efforts to understand the nature of quantum mechanics particularly in realistic models. Here rigorous and precise proofs in the classical domain are presented which put the subject on firmer ground than arguing by analogy with quantum field theory. Moreover, the quantum mechanical and classical calculations aren't completely equivalent because the photon energy is quantized in the quantum theory, but it is not quantized in the classical theory. Thus there is an energy threshold that must be overcome for emission of quantum radiation of a given frequency in QED, but there is no such threshold classically. This may explain why the Dirac current actually leads to radiation classically, but does not in QED. A similar situation arises with the Klein-Gordon current when mixed positive and negative frequency terms are present.

## 2. Classical radiation theory and multipole expansions

Analysis of radiation starts with a current which we will take from quantum wave equations below (we take the currents to have a unit total charge for ease of notation).

$$J^\mu = (c\rho, \mathbf{J}), \quad x^\mu = (ct, \mathbf{x}), \quad g^{00} = +1 \tag{1}$$

The classical electromagnetic field generated by these sources is.

$$\partial_\mu F^{\mu\nu} = \frac{4\pi}{c} J^\nu; \quad F^{\mu\nu} = \partial^\mu A^\nu - \partial^\nu A^\mu \tag{2}$$

$$A^\mu = (\Phi, \mathbf{A}) \tag{3}$$

$$F^{\mu\upsilon} = F^{\mu\upsilon}_{in} + F^{\mu\upsilon}_{ret} = F^{\mu\upsilon}_{out} + F^{\mu\upsilon}_{adv} \tag{4}$$

Where $F^{\mu\upsilon}_{in}$ and $F^{\mu\upsilon}_{out}$ are free fields and where in the Lorentz gauge ($\partial^\mu A_\mu = 0$) we have

$$A^\mu_{ret}(\mathbf{x},t) = \frac{1}{c}\int \frac{J^\mu(\mathbf{x}',t-\frac{R}{c})}{R}d^3x'; \quad R=|\mathbf{x}-\mathbf{x}'| \qquad (5)$$

$$A^\mu_{adv}(\mathbf{x},t) = \frac{1}{c}\int \frac{J^\mu(\mathbf{x}',t+\frac{R}{c})}{R}d^3x'; \quad R=|\mathbf{x}-\mathbf{x}'| \qquad (6)$$

**Definition** A non-radiating source is a 4-current $J^\mu(x)$ for which the integrated Poynting vector over a spherical surface S of radius $R$ vanishes in the limit of large radius for all time using only the retarded fields due to it.

$$\lim_{R\to\infty} \frac{4\pi}{c} \iint_S \mathbf{E}_{ret}(x,t+R/c) \times \mathbf{B}_{ret}(x,t+R/c) \cdot \mathbf{dS} = 0 \text{ for all t.} \qquad (7)$$

## 3. Properties of non-radiating sources

Let $J^\mu(x)$ be a non-radiating source. Any inhomogeneous Lorentz transform of it is also non-radiating as follows from the Lorentz covariance of the Maxwell equations

$$J'^\mu(x) = \Lambda^\mu_{\ \nu} J^\nu(x'), \quad x'^\mu = \Lambda^\mu_{\ \nu} x^\nu + a^\mu \qquad (8)$$

It is also easy to verify that a dilation preserves the non-radiating property

$$J'^\mu(x) = J^\nu(\lambda x), \quad \lambda \text{ a scalar constant} \qquad (9)$$

For many non-radiating sources both the advanced field and the retarded fields are non-radiating (for all time), and since

$$\partial_\mu \left( F^{\mu\nu}_{adv} - F^{\mu\nu}_{ret} \right) = 0 \qquad (10)$$

and since both of these fields would have to fall off more quickly than 1/R at infinity, it follows that they must be exactly equal

$$F^{\mu\upsilon}_{adv} = F^{\mu\upsilon}_{ret} \tag{11}$$

For example, if the source has compact support in space then this is quite generally the case [30]. But when the source does not have compact support, there appears to be no proof that would apply. In the non-radiating cases we study below the advanced and retarded potentials will be found to be the same.

The free Maxwell's equations are invariant under the 15-parameter special conformal group [31], and it is therefore tempting to ask if the non-radiating property is preserved by the elements of this group which includes uniform accelerations. For point particles the consensus until recently has been that a uniformly accelerating charge will radiate energy when viewed in a fixed Lorentz frame [32, 33], but a recent analysis by Singal [34] came to the opposite conclusion. Parrott [33] has argued that the singular nature of the Coulumb field casts some mathematical doubt over Singal's proof. But the arguments of Singal if applied to the present non-radiating sources would seem to be more rigorous since they have no singularities. Thus there is a chance that the quantum wave equations' currents will not radiate if they are undergoing uniform acceleration for all time. If this were found to be the case then all special (15 parameter) conformal transformations of a given non-radiating source would also be non-radiating since any such transformation can be constructed from elements of the Poincare group combined with uniform accelerations and dilations, all of which then preserve the non-radiating property. If this turned out to be true then we would have a new and rather general application of conformal symmetry to nature, and one that does not require (or even allow) the mass of the particle to be zero.

Finally, it is obvious due to linear superposition that the sum of two non-radiating currents is also non-radiating

$$J^{\mu}(x) = J_A^{\mu}(x) + J_B^{\mu}(x) \text{ is non-radiating if } J_A^{\mu} \text{ and } J_B^{\mu} \text{ are} \tag{12}$$

## 4. The Schrödinger equation

Consider the free-particle Schrödinger equation.

$$-\frac{\hbar^2}{2m}\Delta\Psi = i\hbar\frac{\partial\Psi}{\partial t} \tag{13}$$

The charge and current densities are

$$\rho(\mathbf{x},t) = q\Psi^*\Psi; \quad \mathbf{J}(\mathbf{x},t) = \frac{q\hbar}{2mi}\{\Psi^*\nabla\Psi - \Psi\nabla\Psi^*\} = \frac{q\hbar}{2mi}\{\Psi^* \overset{\leftrightarrow}{\nabla} \Psi\} \tag{14}$$

$$\tilde{\Psi}(\mathbf{p},t) = \int \Psi(\mathbf{x},t)e^{-i\mathbf{p}\cdot\mathbf{x}/\hbar}d^3x, \quad \Psi(\mathbf{x},t) = \frac{1}{(2\pi\hbar)^3}\int \tilde{\Psi}(\mathbf{p},t)e^{i\mathbf{p}\cdot\mathbf{x}/\hbar}d^3p \tag{15}$$

**Theorem 1**

Let the wave function $\tilde{\Psi}(\mathbf{p},0)$ at some initial time be in the Schwartz class $S(R^3)$ so that it is everywhere infinitely differentiable and falls off faster than any power, and moreover let it have compact support such that

$$\tilde{\Psi}(\mathbf{p},0) = 0, \text{ for } |\mathbf{p}| > p_{\max} = (1-\varepsilon)mc \text{ for some } \varepsilon>0, \tilde{\Psi}(\mathbf{p},0) \in S(R^3) \tag{16}$$

This condition insures that both the phase velocity $|\mathbf{p}|/2m$ and the group velocity $|\mathbf{p}|/m$ are less than the speed of light.

Then the current generated from the Schrödinger wave is a non-radiating source

**Proof**

The solution for the Schrödinger wave in momentum space is given by

$$\tilde{\Psi}(\mathbf{p},t) = e^{-iE_p t/\hbar}\tilde{\Psi}(\mathbf{p},0), \quad E_p = \mathbf{p}^2/(2m) \tag{17}$$

Therefore it follows that (16) is satisfied for all time. The time dependence can therefore be expanded in a Taylor's series which is convergent for all time t

$$\tilde{\Psi}(\mathbf{p},t) = \sum_{n=0}^{\infty} \frac{\left[-iE_p t/\hbar\right]^n}{n!} \tilde{\Psi}(\mathbf{p},0) \tag{18}$$

It follows that $|\tilde{\Psi}(\mathbf{p},t)| \le \sup_{\mathbf{p}} |\tilde{\Psi}(\mathbf{p},0)| = M < \infty$ since $\tilde{\Psi}(\mathbf{p},0) \in S(R^3)$, and therefore by the Weierstrass M test the inverse Fourier transform of this infinite series is uniformly convergent for all **p**. Therefore it follows that $\Psi(\mathbf{x},t)$ also has a Taylor series expansion in t whose domain of convergence is the entire t axis.

The retarded magnetic field is derived from the vector potential by

$$\mathbf{B}_{ret}(\mathbf{x},t) = \nabla \times \mathbf{A}_{ret} = \nabla \times \frac{1}{c}\int \frac{\mathbf{J}(\mathbf{x}',t-\frac{R}{c})}{R} d^3x' \text{ where } R = |\mathbf{x}-\mathbf{x}'| \tag{19}$$

The Fourier transform of a Schwartz function is also a Schwartz function. Therefore $\Psi(x,t) \in S(R^3)$ and $J_i(x,t) \in S(R^3)$ for all t. Define a unit vector

$$\hat{\mathbf{n}} = \frac{\mathbf{x}-\mathbf{x}'}{|\mathbf{x}-\mathbf{x}'|} \tag{20}$$

The curl can now be evaluated as follows

$$\mathbf{B}_{ret}(\mathbf{x},t) = \frac{1}{c}\int \hat{\mathbf{n}} \times \frac{\partial}{\partial R} \frac{\mathbf{J}(\mathbf{x}',t-\frac{R}{c})}{R} d^3x' \tag{21}$$

$$\mathbf{B}_{ret}(\mathbf{x},t) = \frac{1}{c}\int \hat{\mathbf{n}} \times \left[-\frac{1}{R^2}\mathbf{J}(\mathbf{x}',t-R/c) - \frac{1}{Rc}\frac{\partial}{\partial t}\mathbf{J}(\mathbf{x}',t-R/c)\right] d^3x' \tag{22}$$

In evaluating the radiation emitted, the limit where $|\mathbf{x}| \to \infty$ is taken, and therefore the leading behavior of $\mathbf{B}_{ret}$ is all that need be kept.

$$R = \sqrt{\mathbf{x}^2 + \mathbf{x}'^2 - 2\mathbf{x}\cdot\mathbf{x}'} \tag{23}$$

Define

$$R_0 = |\mathbf{x}| \tag{24}$$

And so to leading order in $R_0$

$$\hat{\mathbf{n}} = \frac{\mathbf{x}}{|\mathbf{x}|} + o(1) \tag{25}$$

$$R = R_0 - \hat{\mathbf{n}} \cdot \mathbf{x'} + o(1) \tag{26}$$

We are interested in keeping only the leading order behavior for large $R_0$, but care must be taken with the time variable. Let us define

$$t' = t - \frac{R_0}{c} \tag{27}$$

And so to leading order in $R_0$

$$\mathbf{B}_{ret}(\mathbf{x}, t' + \frac{R_0}{c}) = -\hat{\mathbf{n}} \times \frac{1}{c^2 R_0} \int \frac{\partial}{\partial t'} \mathbf{J}(\mathbf{x'}, t' + \frac{\hat{\mathbf{n}} \cdot \mathbf{x'}}{c}) d^3 x' + o\left(\frac{1}{R_0}\right) \tag{28}$$

Now expand in a Taylor series in the time variable of $\mathbf{J}$ whose convergence is assured.

$$\mathbf{B}_{ret}(\mathbf{x}, t' + \frac{R_0}{c}) = -\hat{\mathbf{n}} \times \frac{1}{c^2 R_0} \int \left[ \sum_{m=1}^{\infty} \left( \frac{\partial^j}{\partial t'^j} \mathbf{J}(\mathbf{x'}, t') \right) \left( \frac{\hat{\mathbf{n}} \cdot \mathbf{x'}}{c} \right)^{j-1} /(j-1)! \right] d^3 x' + o\left(\frac{1}{R_0}\right) \tag{29}$$

It follows using (16,17) and the Cauchy-Schwarz inequality inequality that

$$\left| \frac{\partial^j}{\partial t^j} \Psi(x,t) \right| \leq \sup_{\mathbf{p}}\left(|\tilde{\Psi}(\mathbf{p},0)|\right) \left( \frac{p_{max}^2}{2m\hbar} \right)^j \left( \frac{4}{3} \pi p_{max}^3 \right) /(2\pi\hbar)^3 \equiv K(j) \tag{30}$$

And

$$\left|\frac{\partial^j}{\partial t^j}\mathbf{J}(x,t)\right| \leq 2^{j+1} K(0)\frac{p_{\max}}{\hbar} K(j) \tag{31}$$

Therefore the infinite series is uniformly convergent again by the Weierstrass M test and the order of integration and summation can be interchanged. One must evaluate the following integrals

$$\mathbf{I}_j(t') = \int \mathbf{J}(\mathbf{x}',t')\left(\frac{\hat{\mathbf{n}}\cdot\mathbf{x}'}{c}\right)^{j-1} d^3x' \tag{32}$$

In terms of which **B** may be written

$$\mathbf{B}_{ret}(x,t'+\frac{R_0}{c}) = -\hat{\mathbf{n}}\times\frac{1}{c^2 R_0}\sum_{j=1}^{\infty}\frac{\partial^j}{\partial t'^j}\mathbf{I}_j(t')\left(\frac{1}{c}\right)^{j-1}/(j-1)! + o\left(\frac{1}{R_0}\right) \tag{33}$$

$\mathbf{I}_j$ takes the form

$$\mathbf{I}_j(t') =$$
$$\frac{q\hbar}{2mi}\int\{\Psi^*(\mathbf{x}',t')\nabla\Psi(\mathbf{x}',t') - \Psi(\mathbf{x}',t')\nabla\Psi^*(\mathbf{x}',t')\}(\hat{\mathbf{n}}\cdot\mathbf{x}')^{j-1} d^3x' \tag{34}$$

this may be written after integration by parts as

$$\mathbf{I}_j(t') = \frac{q}{2m}\int\Psi^*(\mathbf{x}',t')\{\mathbf{p}(\hat{\mathbf{n}}\cdot\mathbf{x}')^{j-1} + (\hat{\mathbf{n}}\cdot\mathbf{x}')^{j-1}\mathbf{p}\}\Psi(\mathbf{x}',t')d^3x' \tag{35}$$

where

$$\mathbf{p} = -i\hbar\nabla \tag{36}$$

Transform (35) by switching to the Heisenberg representation. The time-dependent Heisenberg operators are

$$\mathbf{p}(t) = \mathbf{p}, \quad \mathbf{x}(t) = \mathbf{x} + t\mathbf{p}/m \tag{37}$$

And the quantum expectations $\mathbf{I}_j$ are then

$$\mathbf{I}_j(t') = $$
$$\frac{q}{2m}\int \Psi^*(\mathbf{x},0)\left\{\mathbf{p}(t')(\hat{\mathbf{n}}\cdot\mathbf{x}'(t'))^{j-1} + (\hat{\mathbf{n}}\cdot\mathbf{x}'(t'))^{j-1}\mathbf{p}(t')\right\}\Psi(\mathbf{x},0)d^3x' \qquad (38)$$

$$\mathbf{I}_j(t') = \text{Polynomial of order (j-1) in the variable t'} \qquad (39)$$

Therefore it follows that

$$\frac{\partial^j}{\partial t'^j}\mathbf{I}_j(t') = 0 \qquad (40)$$

And hence to leading order in $R_0$

$$\mathbf{B}_{ret}(\mathbf{x},t'+\frac{R_0}{c}) = o\left(\frac{1}{R_0}\right) \qquad (41)$$

The electric field in the large $R_0$ limit is calculated by

$$\mathbf{E}_{ret}(\mathbf{x},t'+\frac{R_0}{c}) = -\hat{\mathbf{n}}\times\mathbf{B}_{ret}(\mathbf{x},t'+\frac{R_0}{c}) + o\left(\frac{1}{R_0}\right) = o\left(\frac{1}{R_0}\right) \qquad (42)$$

Which follows from the Maxwell equation applied to the far field plane wave limit of the radiation field and is a standard formula in radiation theory.

$$\nabla\times\mathbf{E} + \frac{1}{c}\frac{\partial \mathbf{B}}{\partial t} = 0 \qquad (43)$$

And therefore the integrated Poynting vector over a spherical surface S of radius $R_0$ vanishes in the limit

$$\lim_{R_0 \to \infty} \frac{4\pi}{c} \iint_S \left[ \mathbf{E}_{ret}(\mathbf{x},t'+\frac{R_0}{c}) \times \mathbf{B}_{ret}(\mathbf{x},t'+\frac{R_0}{c}) \right] \cdot \mathbf{dS} = 0 \text{ for all t'} \qquad (44)$$

This completes the proof.

**Corollary** The electromagnetic fields satisfy

$$\lim_{R \to \infty} R F_{ret}^{\mu\nu}(R\hat{\mathbf{x}},t) = 0 \text{ for all t and } \hat{\mathbf{x}} \text{ any unit vector} \qquad (45)$$

Which follows by the same arguments as above but holding the time for the fields fixed instead of the times for the currents.

## 5. The Klein-Gordon free-particle equation

Consider the free complex Klein-Gordon equation.

$$\left[ \partial^\mu \partial_\mu + \frac{m^2 c^2}{\hbar^2} \right] \Phi = 0, \quad m > 0 \qquad (46)$$

The four current we consider is given by

$$J_\mu = \frac{q\hbar i}{2mc^2} \left[ \Phi^* \partial_\mu \Phi - \Phi \partial_\mu \Phi^* \right] \qquad (47)$$

The wave function is in general a superposition of positive and negative frequency terms

$$\Phi = \Phi_+ + \Phi_- \qquad (48)$$

The time evolution equations for these two components are formally given by

$$\Phi_+(\mathbf{x},t') = \exp(-ic\sqrt{\mathbf{p}^2 + m^2 c^2}\, t'/\hbar) \Phi_+(\mathbf{x},0)$$
$$\Phi_-(\mathbf{x},t') = \exp(+ic\sqrt{\mathbf{p}^2 + m^2 c^2}\, t'/\hbar) \Phi_-(\mathbf{x},0) \qquad (49)$$

Where **p** is again given by the canonical form (36). One can make the meaning of the transcendental functions of **p** precise by taking a 3 dimensional Fourier transform and assuming the both functions are members of the Schwartz class so that **p** becomes simply a multiplicative c number. Notice that the radiation is determined completely by the form of the vector current **J** since the far field magnetic field is determined by it alone and the electric field is determined by (42).

**Theorem 2**

Let the wave function again be in the Schwartz class $S(R^3)$ so that it is everywhere infinitely differentiable and falls off faster than any power, and moreover let it have compact support such that

$$\tilde{\Phi}(\mathbf{p},0) = 0, \text{ for } |\mathbf{p}| > p_{max}, \ \tilde{\Psi}(\mathbf{p},0) \in S(R^3), \ p_{max} < \infty \tag{50}$$

This condition insures that the group velocity $|\mathbf{p}|c^2/E_p$ is less than the speed of light (the phase velocity $E_p/|\mathbf{p}|$ is actually greater than c). The time evolution equation ensures that it will be satisfied for all time.

Let the negative frequency part be zero.

Then the current generated from the positive frequency Klein-Gordon wave is a non-radiating source

**Proof**

Everything is as in the Schrödinger case except that now the expression for $\mathbf{I}_j$ becomes

$$\mathbf{I}_j(t') =$$
$$\frac{q\hbar}{2mc^2 i} \int \left\{ \Phi_+^*(\mathbf{x}',t') \nabla \Phi_+(\mathbf{x}',t') - \Phi_+(\mathbf{x}',t') \nabla \Phi_+^*(\mathbf{x}',t') \right\} (\hat{\mathbf{n}} \cdot \mathbf{x}')^{j-1} d^3 x' \tag{51}$$

Again we can introduce a Heisenberg time-dependent position operator in order to evaluate this integral

$$\mathbf{x}_+(t') = \exp(+ic\sqrt{\mathbf{p}^2 + m^2 c^2}\, t'/\hbar) \mathbf{x} \exp(-ic\sqrt{\mathbf{p}^2 + m^2 c^2}\, t'/\hbar)$$
$$= \exp(+ic\sqrt{\mathbf{p}^2 + m^2 c^2}\, t'/\hbar) i\hbar \nabla_p \exp(-ic\sqrt{\mathbf{p}^2 + m^2 c^2}\, t'/\hbar) \tag{52}$$

Which simplifies to

$$\mathbf{x}_+(t') = \mathbf{x} + \frac{\mathbf{p}}{\sqrt{\frac{\mathbf{p}^2}{c^2} + m^2}} t' \tag{53}$$

And once again, because $\mathbf{x}_+(t')$ is linear in $t'$, the $\mathbf{I}_j$ will be polynomials of order (j-1) in $t'$, so that if the wavefunction contains only positive energy terms, there will again be no radiation.

**Corollary**

A Klein-Gordon wave which satisfies all the conditions of Theorem 2 except that it is composed entirely of negative frequency terms produces a current which is a non-radiating source. The proof is a trivial change of the frequency sign in the arguments of Theorem 2.

Let us note that this Heisenberg operator argument is simply a mathematical technique for evaluating the expression for $\mathbf{I}_j$. It does not imply that the position operator so defined has any physical significance. In fact, the position operator for the Klein-Gordon equation is rather a complicated and somewhat controversial subject [28, 29, 35, 36]. Despite this fact, the current that we have used here for the Klein-Gordon equation is the same one that is used in standard field theory analysis of scalar particles where the localizable wave functions are of less importance. In any event, we see that a non-radiating current is naturally arising for the Klein-Gordon equation.

When the wave function contains both positive and negative energy terms, it appears that there will be classical radiation, or at least the previous arguments which lead to zero radiation are no longer valid and there does not seem to be another mechanism to suppress the radiation. Even for mixed wavefunctions however, the radiation at lower frequencies will be suppressed if one considers time averaged currents over a time long compared to $\pi\hbar/(mc^2)$. In this case the cross terms involving + and − wavefunctions in the expression for the current density will average out to be nearly zero.

## 6. The Dirac Equation

The free-particle Dirac equation may be written (using the notation of [37]chapter 1.4).

$$i\hbar \frac{\partial \Psi}{\partial t} = \left(c\boldsymbol{\alpha} \cdot \mathbf{p} + \beta mc^2\right)\Psi, \quad m > 0 \tag{54}$$

The Hamiltonian is given by the operator

$$H = \left(c\boldsymbol{\alpha}\cdot\mathbf{p} + \beta mc^2\right) \tag{55}$$

And the charge and current densities are

$$\rho = q\Psi^{\dagger}\Psi; \quad \mathbf{j} = cq\Psi^{\dagger}\boldsymbol{\alpha}\Psi \tag{56}$$

The time dependent form for the coordinate $\mathbf{x}(t)$ shows oscillatory motion (zitterbewegung). Both H and **p** are independent of time though, and the coordinate satisfies the well-known equation

$$\mathbf{x}(t) = \mathbf{x}(0) + c^2\mathbf{p}H^{-1}t + \frac{\hbar^2}{4}\ddot{\mathbf{x}}(0)H^{-2}(1 - e^{-i2Ht/\hbar}) \tag{57}$$

Where

$$\ddot{\mathbf{x}}(0) = \frac{2c}{i\hbar}(\boldsymbol{\alpha}H - c\mathbf{p}); \quad \dot{\mathbf{x}}(0) = c\boldsymbol{\alpha} \tag{58}$$

Equation (**33**) can be used to calculate B in this case provided the current density for the Dirac equation is used. The expression for $\mathbf{I}_j$ becomes

$$I_j(t') = cq\int \Psi^{\dagger}(\mathbf{x}',t')c\boldsymbol{\alpha}\left(\hat{\mathbf{n}}\cdot\mathbf{x}'\right)^{j-1}\Psi(x',t')d^3x' \tag{59}$$

The operators $\boldsymbol{\alpha}$ and **x** commute, and thus in the Heisenberg representation where they become time dependent, they must also commute at the same time. Therefore, one may shift the time dependence to the operators without regard to their order and obtain

$$I_j(t') = cq\int \Psi^{\dagger}(\mathbf{x}',0)\dot{\mathbf{x}}(t')\left(\hat{\mathbf{n}}\cdot\mathbf{x}'(t')\right)^{j-1}\Psi(x',0)d^3x' \tag{60}$$

It is clear from this expression that all of the $\mathbf{I}_j$ have exponential behavior in time, and therefore the previous arguments used in Theorems 1 and 2 cannot be used to argue that the Dirac equation will not radiate classically. The oscillatory zitterbewegung term has an angular frequency of at least $2mc^2/\hbar$, the same as the oscillation between the cross terms for the Klein-Gordon equation, and so one expects some radiation here at this and higher frequencies. Even solutions with positive energy only will still have this term, and so restricting consideration to them does not solve the problem. For emission of photons at much lower frequencies the oscillatory term can be safely ignored and the classical radiation will be greatly suppressed. Quantum field theorists might be surprised that the radiation does not perfectly vanish classically since the free-particle certainly does not radiate in quantum electrodynamics. The energy quantization condition for the photon in the quantum field theoretical calculation makes it fundamentally different from the classical theory. In classical electromagnetism the zitterbwegung can produce radiation which will not be quantized in photon energies. Thus it can radiate more readily than the quantum counterpart which has an energy gap to overcome before emitting a single photon at the frequency of the zitterbewegung.

We now consider the Foldy-Wouthuysen (FW) transformation [27] and show that a non-radiating current can be easily found if one first makes this transformation, thus eliminating the zitterbewegung problem. The FW transform for a free-particle is a unitary transformation of the form

$$U_{FW} = e^{iS}, \quad S = -(i/2mc)\beta \boldsymbol{\alpha} \cdot \mathbf{p}(mc/p)\tan^{-1}(p/mc) \tag{61}$$

Under this unitary transformation the Hamiltonian operator becomes simply

$$H' = \beta c\left(m^2c^2 + p^2\right)^{1/2} = \beta E_p, \text{ where } \beta = \begin{bmatrix} 1' & 0' \\ 0' & -1' \end{bmatrix} \tag{62}$$

where 0', 1', and -1' are the null and unit matrices of order 2. Breaking the wave function up into a sum a positive and negative energy parts one has

$$\psi' = \phi' + \chi', \quad \phi' = \frac{1+\beta}{2}\psi', \quad \chi' = \frac{1-\beta}{2}\psi' \tag{63}$$

$$i\hbar\frac{\partial \phi'}{\partial t} = E_p \phi' \tag{64}$$

$$i\hbar\frac{\partial \chi'}{\partial t} = -E_p \chi' \tag{65}$$

Both $\phi'$ and $\chi'$ are two component spinors. In [27] it is shown that

$$\psi'(\mathbf{x}) = \int K(\mathbf{x}, \mathbf{x}')\psi(\mathbf{x}')d^3x \tag{66}$$

$$K(\mathbf{x},\mathbf{x'}) = \frac{1}{(2\pi)^3} \int \left[\frac{2E_{p'}}{E_{p'}+mc^2}\right]^{\frac{1}{2}} \left[1 + \frac{\beta(\beta mc + \boldsymbol{\alpha}\cdot\mathbf{p})}{E_{p'}}\right] \exp(i\mathbf{p'}\cdot(\mathbf{x}-\mathbf{x'})) d^3 p' \qquad (67)$$

And so the Foldy Wouthuysen trasformation is an integral transform of the wave function. "In general, $\psi'$ at a given point is constituted from contributions depending on $\psi$ over a neighborhood of dimensions of the order of a Compton wavelength of the particle about the point." The transformed coordinate operator is as follows

$$\mathbf{x'} = e^{iS}\mathbf{x}e^{-iS} = \mathbf{x} - \frac{ic\hbar\beta\boldsymbol{\alpha}}{2E_p} + \frac{i\beta\hbar c^2(\boldsymbol{\alpha}\cdot\mathbf{p})\mathbf{p} - [\boldsymbol{\sigma}\times\mathbf{p}]\hbar c^2 p}{2E_p(E_p+mc^2)p} \qquad (68)$$

The inverse of this is called the "mean position operator".

$$X_{mean} = e^{-iS}\mathbf{x}e^{iS} = \mathbf{x} + \frac{ic\hbar\beta\boldsymbol{\alpha}}{2E_p} - \frac{i\beta\hbar c^2(\boldsymbol{\alpha}\cdot\mathbf{p})\mathbf{p} + [\boldsymbol{\sigma}\times\mathbf{p}]\hbar c^2 p}{2E_p(E_p+mc^2)p} \qquad (69)$$

This same position operator was identified by Newton and Wigner [28, 38] as the most natural operator to identify with the position operator for the Dirac equation because of its localizability properties.

We are now in a position to identify a non-radiating current source. We see that after performing the Foldy-Wouthuysen transformation that both $\phi'$ and $\chi'$ satisfy the Klein-Gordon equation and the simple time evolution equations

$$\phi'(\mathbf{x},t_0) = \exp(-ic\sqrt{\mathbf{p}^2+m^2c^2}\,t_0/\hbar)\phi'(\mathbf{x},0)$$
$$\chi'(\mathbf{x},t_0) = \exp(+ic\sqrt{\mathbf{p}^2+m^2c^2}\,t_0/\hbar)\chi'(\mathbf{x},0) \qquad (70)$$

**Theorem 3**

Consider a positive energy solution to the Dirac equation which has undergone a Foldy-Wouthuysen transformation. Define a conserved current by analogy with the Klein-Gordon current

$$J_\mu^{FW} = \frac{q}{2mi}\phi'^\dagger \overrightarrow{\partial_\mu}\phi' \tag{71}$$

Let the spinor wave function elements be of Schwarz class with compact support in p space as in theorem 2. Then this current is a non-radiating source.

**Proof**

The proof is identical to theorem 2.

So by performing a Foldy-Wouthuysen transformation a non-radiating current can be defined for the Dirac equation, but this is not the usual current density used in quantum electrodynamics.

**Corollary**

The current obtained from a purely negative energy solution to the Dirac equation is also a non-radiating as was the case for the Klein-Gordon field.

The current (71) is not the usual one used in the Foldy-Wouthuysen transformation. There are two conserved quantities here: $\left(\phi'^\dagger \phi'\right)$ and $\left(\phi'^\dagger \dfrac{\overrightarrow{\partial}}{\partial t}\phi'\right)$, and either one can be used to construct a 4-current, but the choice (71) leads to the simplest illustration of a non-radiating 4-current. The usual choice $\left(\phi'^\dagger \phi'\right)$ does not lead to a unique or simple 4-current. Our point here is that one can identify a non-radiating source from the wave equation quite naturally. We do not claim that this current should be the one identified with the probability current of the quantum theory.

## 7. Higher spin wave equations

We follow the approach developed by Foldy [29] which is based on the irreducible representations of the Poincare group as analyzed by Bargmann and Wigner [39]. Foldy shows how a very wide class of relativistic wave equations can be written in a "canonical form" which is analogous to the Foldy-Wouthuysen transformation for spin ½. For a particle of mass m and spin s the "canonical form" is

$$i\hbar\partial\chi(\mathbf{r},t)/\partial t = \beta E_p \chi(\mathbf{r},t) \tag{72}$$

$$E_p = c\left(m^2c^2 + p^2\right)^{1/2} \tag{73}$$

$$\beta = \begin{pmatrix} 1' & 0' \\ 0' & -1' \end{pmatrix} \tag{74}$$

Where $\chi(\mathbf{r},t)$ is a 2(2s+1)-component wave function, b is a 2(2s+1)x2(2s+1) diagonal Hermetian matrix and 0' and 1' are the null and unit matrices of order 2s+1. The top half the elements of $\chi(\mathbf{r},t)$ transform as an irreducible representation of the Lorentz group, as does the lower half. We can immediately write down a non-radiating current expression using the same technique as in the Foldy-Wouthuysen and Klein-Gordon cases.

$$J_\mu = \frac{q\hbar i}{2mc^2}\left[\chi^\dagger \overleftrightarrow{\partial}_\mu \chi\right] \tag{75}$$

We can immediately state

**Theorem 4** Let the elements of $\chi(\mathbf{r},t)$ be belong to the Schwartz class of compact support as in the Klein-Gordon case. Then the current defined by (75) is a non-radiating current.

The same comments about the current chosen for the Foldy-Wouthuysen theory apply as well here. The current (75) is not the one proposed by Foldy as a probability current for the quantum theory.

Note that mixed positive and negative energy states are allowed in this theory by the current definition (75). This theory also applies to the Klein-Gordon equation (which would be described by a two element spinor) and the Foldy-Wouthuysen transformation. By taking the current to be the sum of the positive energy current and the negative energy current as (75) does, then the mixed states are non-radiating for all spins.

## 8. Phase-space wave equations

The shortcomings of the Klein-Gordon, Dirac, and Foldy-Wouthhusen equations as satisfactory relativistic wave equations have been pointed out be Prugovecki [36]. With coworkers he has proposed a wide class of alternative wave equations which overcome these shortcomings [36, 40-42]. For the spinless non-relativistic case the phase-space wave function is written as[36] (page 43, eq. 8.22)

$$\psi(\mathbf{q},\mathbf{p};t) = \int \exp(-i\mathbf{p}\cdot\mathbf{x}/\hbar)\xi(x-q)\psi(x,t)d^3x \tag{76}$$

Where $\xi$ is the resolution generator function and $\psi$ the usual wave function. The currents are given by [36] (page 41, eqs. 8.12 - 8.14)

$$i\hbar\partial_t\psi(\mathbf{q},\mathbf{p};t) = -(\hbar^2/2m)\nabla_\mathbf{q}^2\psi(\mathbf{q},\mathbf{p};t) \tag{77}$$

$$\rho(\mathbf{q},\mathbf{p};t) = |\psi(\mathbf{q},\mathbf{p};t)|^2 \tag{78}$$

$$\mathbf{j}^\xi(\mathbf{q},\mathbf{p};t) = \hbar/(2im)\psi^*(\mathbf{q},\mathbf{p};t)\overset{\leftrightarrow}{\nabla}_\mathbf{q}\psi(\mathbf{q},\mathbf{p};t) \tag{79}$$

$$\partial_t\rho(\mathbf{q},\mathbf{p};t) + \nabla_\mathbf{q}\cdot\mathbf{j}^\xi(\mathbf{q},\mathbf{p};t) = 0 \tag{80}$$

$$\rho(\mathbf{q},t) = \int \rho(\mathbf{q},\mathbf{p};t)d^3p; \quad \mathbf{j}^\xi(\mathbf{q},t) = \int \mathbf{j}^\xi(\mathbf{q},\mathbf{p};t)d^3p \tag{81}$$

If for each $\mathbf{p}$ held fixed the functions $\psi(\mathbf{q},\mathbf{p};t)$ satisfy the regularity conditions of Theorem 1 then each these currents will all be non-radiating. If the integrals in (81) are uniformly convergent, then owing to superposition as in (12), the resulting currents will be non-radiating.

For the spinless relativistic case we have [36] (page 114, eqs. 8.11-8.13)

$$J_\eta^\mu(q) = i\hbar/(mZ_\eta)\int \phi^*(q,p)\overset{\leftrightarrow}{\partial}^\mu\phi(q,p)\delta(p^2-m^2c^2)d^4p \tag{82}$$

Where q and p are 4 vectors here, and where $\phi(q,p)$ satisfies the Klein-Gordon equation in the q variable for p fixed. The parameter $Z_\eta$ is an integration constant. Once again, by superposition this current will be a non-radiating source with the assumptions of Theorem 2 and assuming uniform convergence of the integrals over the momenta in this equation. Generalizing these results to higher spin cases is trivial in the Prugovecki theory since the spin decouples from the orbital motion for free particles in these theories, and so we have a very large class of non-radiating currents generated from Prugovecki's phase space wave equations.

## 9. Advanced Fields

Let us consider the advanced field solutions.

$$\mathbf{B}_{adv}(\mathbf{x},t) = \nabla \times \mathbf{A}_{adv} = \nabla \times \frac{1}{c}\int \frac{\mathbf{J}(\mathbf{x}',t+\frac{R}{c})}{R} d^3x' \text{ where } R = |\mathbf{x}-\mathbf{x}'| \qquad (83)$$

$$\mathbf{B}_{adv}(\mathbf{x},t) = \frac{1}{c}\int \hat{\mathbf{n}} \times \frac{\partial}{\partial R} \frac{\mathbf{J}(\mathbf{x}',t+\frac{R}{c})}{R} d^3x' \qquad (84)$$

$$\mathbf{B}_{adv}(\mathbf{x},t) = \frac{1}{c}\int \hat{\mathbf{n}} \times \left[ -\frac{1}{R^2}\mathbf{J}(\mathbf{x}',t+R/c) + \frac{1}{Rc}\frac{\partial}{\partial t}\mathbf{J}(\mathbf{x}',t+R/c) \right] d^3x' \qquad (85)$$

$$\mathbf{B}_{adv}(\mathbf{x},t'+\frac{R_0}{c}) = \hat{\mathbf{n}} \times \frac{1}{c^2 R_0}\int \frac{\partial}{\partial t'}\mathbf{J}(\mathbf{x}',t'-\frac{\hat{\mathbf{n}}\cdot\mathbf{x}'}{c}) d^3x' + o\left(\frac{1}{R_0}\right) \qquad (86)$$

The same Taylor series convergence arguments apply as in the retarded case, therefore

$$\mathbf{B}_{adv}(x,t'+\frac{R_0}{c}) = \hat{\mathbf{n}} \times \frac{1}{c^2 R_0} \sum_{j=1}^{\infty} \frac{\partial^j}{\partial t'^j}\mathbf{I}_j(t')\left(\frac{-1}{c}\right)^{j-1}/(j-1)! + o\left(\frac{1}{R_0}\right) \qquad (87)$$

But since all the $\mathbf{I}_j$ terms vanished after taking the time derivatives in all the above non-radiating cases, we have

$$\mathbf{B}_{adv}(x,t'+\frac{R_0}{c}) = o\left(\frac{1}{R_0}\right) \text{ and consequently } \mathbf{E}_{adv}(x,t'+\frac{R_0}{c}) = o\left(\frac{1}{R_0}\right) \qquad (88)$$

It also follows that

$$\lim_{R\to\infty} RF_{adv}^{\mu\nu}(R\hat{\mathbf{x}},t) = 0 \text{ for all t and } \hat{\mathbf{x}} \text{ any unit vector} \qquad (89)$$

**Theorem 5** For any non-radiating current derived from wave equations above, it follows that both the advanced and retarded fields fall off faster than 1/R and since

$$\Box\left(F_{adv}^{\mu\upsilon}(\mathbf{x},t) - F_{ret}^{\mu\upsilon}(\mathbf{x},t)\right) = 0, \text{ for all } \mathbf{x} \text{ and t.} \tag{90}$$

and the only solution to this equation that falls off faster than 1/R is zero and so

$$F_{adv}^{\mu\upsilon}(\mathbf{x},t) = F_{ret}^{\mu\upsilon}(\mathbf{x},t), \text{ for all } \mathbf{x} \text{ and t.} \tag{91}$$

## 10. Neutral Particles

Neutral particles will generally have some intrinsic non-vanishing magnetic or electric moments of higher order. The free-particle quantum wave equations are non-radiating for these objects as well if the electromagnetic fields are treated classically provided the various charge and current moments are independent time and of the particle's position. In this case, as all moments can be thought of as superpositions of electric and magnetic charges, and owing to the symmetry between magnetic and electric charges in Maxwell's equations, and the fact that the sum of two or more non-radiating currents will also be non-radiating, it can be concluded that there will not be any radiation from any electromagnetic higher moment.

## 11. A possible explanation of why a non-linear term in quantum mechanical wave equations has not been observed.

The possibility of nonlinear corrections to Schrödinger's equation have long been considered [43, 44]. There is no experimental evidence for them [23]. If present they would invalidate the non-radiating proofs of the Theorems given above. Perhaps the reason why they have not been observed is they would lead to electromagnetic radiation emitted from a free charged particle when the radiation field is treated classically. They would also lead to radiation for neutral particles which had some intrinsic nonzero electromagnetic moment. Undoubtedly, if quantum electrodynamics could be extended to include such nonlinear terms, the radiation would be nonzero there as well even for free particles. We note that there are some solutions of nonlinear wave equations which would not radiate, such as the soliton solutions found in [43] with a logarithmic potential. But these are special cases. Most solutions would radiate, even for the log potential, although radiation would be a mechanism for the system to relax into a soliton equilibrium state (if such a solution existed).

## 12. A stochastic interpretation of the null radiation result

Consider a random classical radiation field. Assume that the equation of motion is such that the incoming field averages to zero and the presence of the particle does not produce any net radiation on the average.

$F^{\mu\nu}_{vacuum}(x, t)$ is a random electromagnetic field satisfying

$$E(F^{\mu\nu}_{vacuum}(x, t)) = 0 \tag{92}$$

Now imagine placing a charged particle in this vacuum which will be accelerated by the electromagnetic force and possibly other forces and will also modify the vacuum fields

$$F^{\mu\nu}(x, t) = F^{\mu\nu}_{vacuum}(x, t) + F^{\mu\nu}_{rad}(x, t) \tag{93}$$

$$F^{\mu\nu}_{rad}(x, t) = \partial^\mu A^\nu_{rad} - \partial^\nu A^\mu_{rad}; \quad A^\mu_{rad}(x, t) = \frac{1}{c}\int \frac{J^\mu(x', t - \frac{R}{c})}{R} d^3x'; \quad R = |x - x'| \tag{94}$$

If the particle is in equilibrium with the vacuum, it is necessary that it doesn't radiate on the average. Equilibrium will of course depend on the details of the radiation field, and the equation of motion of the particle, but without specifying these, we can still learn something. If we take an ensemble average, the incoming wave averages to zero and it follows that.

$$E(F^{\mu\nu}_{rad}(x, t)) = o\left(\frac{1}{R}\right) \text{ and so therefore } E(J^\mu(x', t - \frac{R}{c})) \text{ is non-radiating} \tag{95}$$

This is just what we have found for various free-particle quantum wave equations. This radiationless behavior is a natural relaxation mechanism that can be understood qualitatively in a stochastic model.

## 13. Conclusion

There is a deep connection between the theory of non-radiating sources and quantum mechanics. It is interesting to speculate that in the presence of forces a new variational principle based on minimizing the electromagnetic radiation might be found to provide an alternative derivation of quantum wave equations.

The usual Dirac current does not satisfy the classical non-radiating condition. But by making a Foldy-Wouthuysen transformation it is possible to identify a non-radiating current. This might contradict the intuition of field theorists. It is helpful to keep in mind that the radiation of the Dirac current, due as it is to the Zitterbewegung term, would correspond to a photon with the energy of twice the electron rest mass in the quantum theory, and so it could not contribute in the soft-photon limit of bremsstrahlung. But in the classical field theory any frequency can contribute in the soft-photon limit since there is no quantization of photon energy. This explains why the free Dirac particle appears to radiate in the classical theory whereas it certainly does not in the quantum field theory. The same comments apply to the Klein-Gordon case in mixed positive and negative frequency states.

In Foldy's canonical form for higher spin wave equations a non-radiating current can be defined for all cases as long as the mass is nonzero. Even mixed positive and energy states are non-radiating in these cases. Progovecki's phase-space wave equations also produce currents which are non-radiating.

We have seen that for the wave equations the advanced and retarded fields are the same. This means that there is no preferred arrow of time for classical electromagentism applied to these systems.

The lack of evidence for a nonlinear term in quantum mechanics may be a result of the fact that such a nonlinear term would cause the free-particle system to radiate.

If we accept that uniformly accelerating charges do not radiate, then any special conformal transformation applied to a non-radiating 4-current yields another non-radiating 4-current. This leads to a new application for conformal symmetry which applies to massive wave equations and particles. Their currents are a subset of a conformally invariant set of currents all of which are non-radiating.

The rich class of non-radiating 4-current solutions presented here may have practical applications in designing such systems, and may help in the search for their most general mathematical form.

## Acknowledgements

The author wishes to thank Greg Gbur for helpful comments and Iwo Bialynicki-Birula for an interesting and insightful discussion.

## References

[1]  A. Sommerfeld, Akad. der Wiss. (Gött.), Math. Phys. Klasse, Nach., 363 (1904).

[2]  A. Sommerfeld, Akad. der Wiss. (Gött.), Math. Phys. Klasse, Nach., 99 (1904).

[3]  G. A. Schott, Proc. R. Soc. London, Ser. A, **159**, 570–591 (1937).

[4]  G. A. Schott, Phil. Mag., **15**, 752–761 (1933).

[5]     D. Bohm and M. Weinstein, Phys. Rev. **74**, 1789–1798 (1948).

[6]     G. H. Goedecke, Phys. Rev. B **135**, 281 (1964).

[7]     T. Waite, A. O. Barut, and J. R. Zeni, in *Electron Theory and Quantum Electrodynamics: 100 Years Later (Nato Asi Series. Series B, Physics, Vol 358)*, edited by J. Dowling (Plenum Press, London, 1997).

[8]     P. Pearl, in *Electromagnetism: Paths to Research*, edited by D. Teplitz (Plenum Press, New York, 1982), p. 211–295.

[9]     G. Gbur, Progress in Optics **45**, 273 (2003).

[10]    G. Gbur, in *Ph. D. Thesis, Department of Physics and Astronomy* (University of Rochester, Rochester, 2001).

[11]    W. C. Chew, Y. M. Wang, G. Otto, et al., lnverse Problems **10**, 547 (1994).

[12]    B. J. Hoenders, J. Opt. Soc. Am. A **14**, 262 (1997).

[13]    A. Carati and L. Galgani, Nuovo Cimento **118 B**, 839 (2003).

[14]    N. Bleistein and J. K. Cohen, J. Math. Phys., **18**, 194 (1977).

[15]    E. A. Marengo and R. W. Ziolkowski, Radio Science **37** (2002).

[16]    A. J. Devaney, J. Opt. Soc. Am. A **21**, 2216 (2004).

[17]    A. J. Devaney and R. P. Porter, J. Opt. Soc. Am. A **2**, 2006 (1985).

[18]    A. Gamliel, K. Kim, A. I. Nachman, et al., J. Opt. Soc. Am. A **6**, 1388 (1989).

[19]    A. J. Devaney and R. P. Porter, J. Opt. Soc. Am. **72**, 327 (1982).

[20]    E. A. Marengo and R. W. Ziolkowski, Phys. Rev. Lett. **83**, 3345 (1999).

[21]    E. A. Marengo, A. J. Devaney, and R. W. Ziolkowski, J. Opt. Soc. Am. A **16**, 1612 (1999).

[22]    M. P. Davidson, Annales Fondation Louis de Broglie **30**, 259 (2005).

[23]    R. Gahler, A. G. Klein, and A. Zeilinger, Phys. Rev. A **23**, 1611 (1981).

[24]    J. A. Wheeler and R. P. Feynman, Rev. Mod. Phys. **17**, 157 (1945).

[25]    J. A. Wheeler and R. P. Feynman, Rev. Mod. Phys. **21**, 425 (1949).

[26]    G. N. Plass, in *Ph. D. Thesis Department of Physics* (Princeton University, Princeton, 1947).

[27]    L. L. Foldy and S. A. Wouthuysen, Phys. Rev. **78**, 29 (1950).

[28]    T. D. Newton and E. P. Wigner, Rev. Mod. Phys. **21**, 400 (1949).

[29]    L. L. Foldy, Phys. Rev. **102**, 568 (1956).

[30]    F. G. Friedlander, Proceedings of the London Mathematical Society **27**, 551 (1973).

[31]    T. Fulton, F. Rohrlich, and L. Witten, Rev. Mod. Phys. **34**, 444 (1962).

[32]    F. Rohrlich, *Classical Charged Particles* (Addison-Wesley, Reading, 1965).

[33] S. Parrott, www.arxiv.org/abs/gr-qc/9711027 (2006).

[34] A. Singal, General Relativity and Gravitation **27**, 953 (1995).

[35] H. Feshbach and F. Villars, Rev. Mod. Phys. **30**, 24 (1958).

[36] E. Prugovecki, *Stochastic Quantum Mechanics and Quantum Spacetime* (D. Reidel, Dordrecht, 1984).

[37] J. D. Bjorken and S. D. Drell, *Relativistic Quantum Mechanics* (McGraw-Hill, New York, 1964).

[38] A. S. Wightman, Rev. Mod. Phys., 845 (1962).

[39] V. Bargmann and E. P. Wigner, Proc. Nat. Acad. Sci. U. S. **34**, 211 (1948).

[40] F. E. Schroeck, *Quantum mechanics on phase space* (Kluwer Academic, Dordrecht, 1996).

[41] S. T. Ali, J. A. Brooke, P. Busch, et al., Can. J. Phys. **66**, 238 (1988).

[42] S. T. Ali, R. Gagnon, and E. Prugovecki, Can. J. Phys., **59**, 807 (1981).

[43] I. Bialynicki-Birula and J. Mycielski, Ann. Phys. **100**, 62 (1976).

[44] S. Weinberg, Phys. Rev. Lett. **62**, 485 (1989).